\documentclass[10pt,conference]{IEEEtran}

\usepackage{cite}
\usepackage{amsmath}
\usepackage{graphicx}
\usepackage{booktabs}
\usepackage{array}
\usepackage{xcolor}
\usepackage{listings}
\usepackage[hidelinks]{hyperref}

\newcommand{\repourl}{\url{https://github.com/higolab/SZZBench}}
\newcommand{\releasedoi}{\url{https://doi.org/10.5281/zenodo.23685149}}
\newcommand{\tool}{SZZBench}
\newcommand{\cmd}[1]{\texttt{#1}}

\begin{document}

\title{SZZBench: Controlled Git Histories with Explicit Oracles for Diagnosing SZZ Implementations}

\author{
\IEEEauthorblockN{Yoshiki Higo}
\IEEEauthorblockA{The University of Osaka\\Japan\\higo@ist.osaka-u.ac.jp}
\and
\IEEEauthorblockN{Inase Kondo}
\IEEEauthorblockA{The University of Osaka\\Japan\\i-kondou@ist.osaka-u.ac.jp}
\and
\IEEEauthorblockN{Masanari Kondo}
\IEEEauthorblockA{Kyushu University\\Japan\\kondo@ait.kyushu-u.ac.jp}
\and
\IEEEauthorblockN{Yasutaka Kamei}
\IEEEauthorblockA{Kyushu University\\Japan\\kamei@ait.kyushu-u.ac.jp}
}

\maketitle

\begin{abstract}
The SZZ algorithm traces a bug-fixing commit back to the commits that introduced the bug. Its labels
underlie defect-prediction data sets and many mining software repositories studies. SZZ implementations are mostly evaluated against oracles
mined from real projects. Such oracles are realistic, but each case mixes the history factors that make
an implementation succeed or fail, and an aggregate score cannot show why. We present \tool{}, a
tool that shows why: it generates ordinary Git repositories from declarative
scenarios with explicit oracles, scores the predictions of any SZZ implementation under declared
evaluation policies, and reads each result against 26 naive baselines, such as always answering the
second commit. \tool{} also checks the declarations of new scenarios, such as the tier, against the results of the naive baselines. Currently, \tool{} ships 279
scenarios in 17 categories and adapters for four SZZ tools. Running these tools in nine configurations
without changing their code, we find that PyDriller passes 161 scenarios and a rule that always answers
the second commit passes 160, yet only 97 scenarios are passed by both: the two succeed on different
scenarios, which the number of passed scenarios hides and the per-scenario results reveal. \tool{} is
available at \repourl.
\end{abstract}


\section{Introduction and Positioning}\label{sec:intro}

The SZZ algorithm~\cite{Sliwerski2005When} starts from a bug-fixing commit (BFC), blames the lines the
fix changes, and reports the commits that last touched them as bug-inducing commits (BICs). Its labels
serve as the ground truth of just-in-time defect prediction~\cite{Kamei2013JITQA} and of many empirical
studies in mining software repositories~\cite{RodriguezPerez2018Reproducibility}. Many
implementations and variants exist, from annotation graphs~\cite{Kim2006Automatic}, meta-change
filtering~\cite{DaCosta2017Framework}, candidate selection~\cite{Davies2014TextDependence}, and
refactoring awareness~\cite{CamposNeto2018RefactoringImpact} to token-level
histories~\cite{German2019Cregit,Kondo2026MVSZZ}, learned models~\cite{Tang2023NeuralSZZ}, LLMs~\cite{Tang2025LLM4SZZ},
and LLM-based agents~\cite{Risse2026LLMAgents}. The choice matters: mislabeled changes
affect defect prediction~\cite{Fan2021Mislabeled}, and the SZZ variant changes the results of
vulnerability prediction~\cite{Cannavale2025GroundTruth}.

Evaluating an SZZ implementation requires knowing which commits truly induced a bug. Oracles mined
from real projects, such as developer-informed oracles~\cite{Rosa2021DeveloperInformed,Rosa2023Comprehensive}
and the \cmd{Fixes:} tags of the Linux kernel~\cite{Lyu2024LinuxSZZ}, show how well an implementation
finds real BICs. Each of their cases, however, mixes several history factors at once, such as
intermediate edits, refactorings, merges, and fixes that only add lines, so whether an implementation
identifies a BIC correctly does not tell which conditions it handles or why it succeeds or fails.
Consequently, we expected controlled, artificial histories to make these causes easier to identify.

\textbf{Related work.} The closest work is that of \c{S}erifo\u{g}lu and
T\"uz\"un~\cite{Serifoglu2025CornerCases}, who hand-crafted one repository with corner-case bugs and
found that SZZ and two LLMs did worse on these bugs than on ordinary ones. \tool{} turns that idea into
a tool: it generates ordinary Git repositories from scenarios declared in YAML text files; it makes the
oracle and the evaluation policy explicit and scores every SZZ implementation under the same
conditions; it ships naive baselines, such as always answering the second commit, against which results
are read and scenarios are checked; and it also helps its users write new scenarios. \tool{} does not replace real-project oracles, which are needed to estimate accuracy in
practice; it isolates the conditions behind a success or a failure. Benchmarks with known ground truth
serve the same diagnostic role in other fields, such as injected bugs for bug
finders~\cite{DolanGavitt2016LAVA} and hand-built oracles for tracking code
histories~\cite{Jodavi2022CodeTracker}. Recent agent-based work reports that many BICs lie beyond what
\cmd{git blame} returns~\cite{Shi2026AgenticSZZ}; \tool{} also contains scenarios of that kind by design
(Section~\ref{sec:floor}).

To show why an SZZ implementation succeeds or fails, this paper makes three contributions:
\begin{enumerate}
\item a tool that generates controlled, reproducible Git histories with explicit oracles and evaluation
  policies, and scores SZZ tools against them (Section~\ref{sec:design});
\item built-in naive baselines and layered checks that support diagnosis and the quality control of
  scenarios (Sections~\ref{sec:floor} and~\ref{sec:suite});
\item a reusable suite of 279 scenarios in 17 categories, adapters for four SZZ tools, and recorded runs
  of nine configurations (Sections~\ref{sec:suite} and~\ref{sec:eval}).
\end{enumerate}
{\raggedright
\tool{} is available at \repourl{}, and the release is archived at \releasedoi.\par}

\begin{figure*}[t]
\centering
\includegraphics[width=\textwidth]{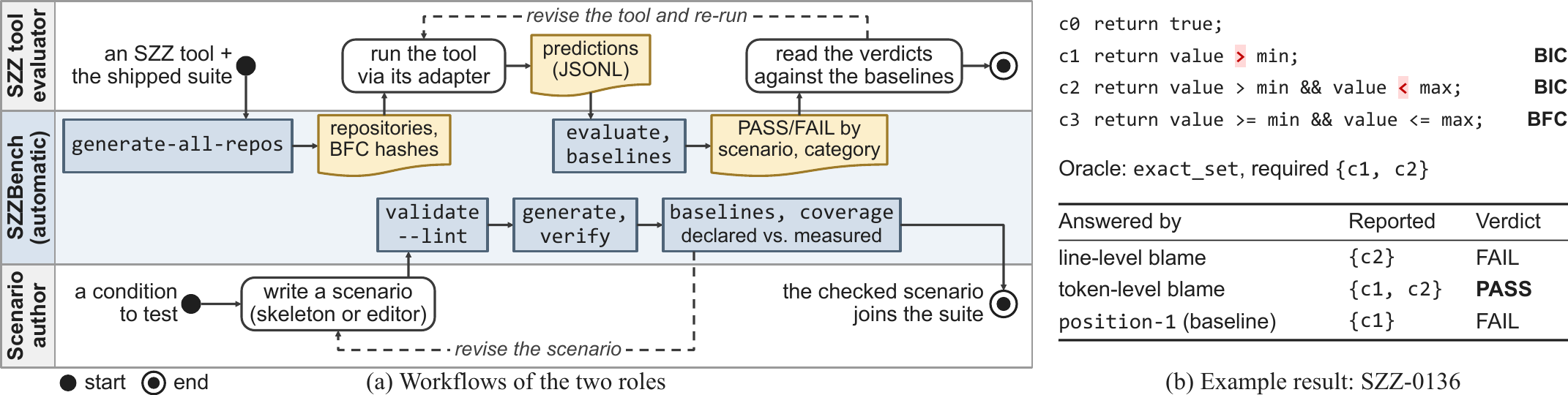}
\caption{(a) An evaluator starts from an SZZ tool and the shipped suite; an author's flow ends when a checked
scenario joins that suite. People do the steps in the outer lanes, and those in the middle lane run
automatically. (b) Two commits write the faulty condition (red), and one fix repairs both bounds.}
\label{fig:workflow}
\end{figure*}

\section{\tool{} Design and Workflow}\label{sec:design}

Fig.~\ref{fig:workflow}(a) shows the workflows of the two user roles, the SZZ tool evaluator and the
scenario author, and Fig.~\ref{fig:workflow}(b) shows a result for SZZ-0136, one of the scenarios shipped
with \tool{}. \tool{} works in three steps. \cmd{generate-all-repos} replays each scenario into an
ordinary Git repository. An adapter runs an SZZ tool on it and writes the BICs the tool reports, or its
error, as one record per scenario in a JSON Lines predictions file. For SZZ-0136, line-level blame
(Fig.~\ref{fig:workflow}(b)) yields this record:
\begin{lstlisting}
{"scenario_id": "SZZ-0136",
 "repository_id": "29a5a1f7990b88e9",
 "predictions": [{"fix_commit": "ecaab90...",
   "inducing_commit": "09c2d7d..."}]}
\end{lstlisting}
\noindent\cmd{evaluate} then scores each record PASS or FAIL against the scenario's oracle: the BICs a
correct answer must report.
The JSONL interface separates SZZ-specific execution from generation and scoring, allowing new SZZ
implementations to reuse the same generator, evaluator, and baselines.

\subsection{Scenarios and Generated Repositories}

A scenario is a YAML file that lists logical commits (\cmd{c0}, \cmd{c1}, \ldots) with their authors,
dates, roles, and file operations, together with the bugs that are introduced and fixed in those
commits. Classic scenarios describe a history on a single branch; graph scenarios also describe
branches and merges, cherry-picks, reverts, and rebases, as well as file renames and copies.
\cmd{generate-all-repos} writes two metadata files next to each repository: \cmd{commit-map.json} maps
logical ids to commit hashes, and \cmd{manifest.json} identifies the generated repository by a
\cmd{repository\_id}. An SZZ tool receives only the repository and the hash of each BFC.

\textbf{Reproducibility.} Generating a repository is deterministic and isolated from the user's Git
configuration: the same scenario yields the same commit hashes, so predictions recorded once can be
scored again later without rerunning the SZZ tool.

\subsection{Oracles and Evaluation Policies}

In SZZ-0136 (Fig.~\ref{fig:workflow}(b)), \cmd{c1} and \cmd{c2} each wrote part of the faulty condition,
so its oracle names both as BICs under the policy \cmd{exact\_set}: a result passes if and only if it
reports both and nothing else. Blaming the line that the fix deletes reaches only \cmd{c2}, the last commit to touch it,
so line-level blame fails; token-level blame attributes \cmd{>} to \cmd{c1} and \cmd{<} to \cmd{c2} and
passes (Section~\ref{sec:eval}). Four other policies relax or reshape this rule: \cmd{partial\_credit}
passes with a minimum number of required BICs and no false positive, \cmd{top\_k} scores a ranking, and
\cmd{relaxed\_with\_related} and \cmd{relaxed\_with\_credit} tolerate declared related commits. An oracle
can also demand an explicit empty answer (\cmd{expect\_no\_bic}).
Oracles distinguish the \emph{direct} BIC, which brought a fault into the code that the fix repairs,
from its \emph{origin}, for example the commit that first wrote a line that was later copied; the
policy states which one counts. A required BIC need not fail by itself: the oracle requires every
commit whose change is necessary for the failure that the BFC repairs, the commit at which the failure
first appears and the earlier commits it depends on. For a scenario with several BFCs, the reported
BICs must match each BFC's own BICs. \cmd{evaluate} also scores a scenario FAIL when its record reports
an SZZ tool's error, when the record is missing, or when the predictions were made for an older version
of the generated repository, so that a crash is never mistaken for an empty answer. Of the 279
scenarios, 267 use \cmd{exact\_set}.

\subsection{Naive Baselines and Tiers}\label{sec:floor}

\tool{} scores 26 naive baselines with \cmd{evaluate}, the command that scores an SZZ tool. They form
four families:
\begin{itemize}
\item \emph{metadata}: one rule per author, which reports that author's commits, and
  \cmd{message-admits-fault}, which reports the commits whose message admits a fault;
\item {\raggedright \emph{position}: \cmd{position-k}, the commit at index $k$ (\cmd{position-1} is the second
  commit), \cmd{position-before-fix}, \cmd{position-all-non-fix}, and \cmd{abstain}, which reports
  nothing;\par}%
\item \emph{shape}: \cmd{shape-last-toucher}, the newest commit to touch each file that the fix
  touches;
\item \emph{blame}: \cmd{blame-deleted-lines}, which blames every line that the fix deletes, i.e., naive
  B-SZZ.
\end{itemize}
A single baseline applied to every scenario sets a floor that an SZZ implementation should beat:
\cmd{position-1} passes 160 scenarios, and \cmd{blame-deleted-lines} 156. The union of all baselines, 233 scenarios, is an upper bound obtained by choosing the best
baseline for each scenario; it is not the score of any one predictor.

Every scenario declares a \emph{tier}. A \emph{control} scenario checks that an SZZ tool handles a
condition that some naive baseline may also solve; a \emph{discriminating} scenario is designed so that
no declared baseline solves it, and \cmd{baselines} checks the declaration. The tier is thus
relative to the declared baselines and is not a general measure of difficulty. A scenario whose oracle
has one BIC is always solved by some \cmd{position-k}, and an empty oracle by \cmd{abstain}, so all 46
discriminating scenarios currently shipped require two or three BICs for one BFC. In 35 of the 46, one required BIC was
designed to lie beyond what blaming the fix's deleted lines returns. The results of an SZZ tool on the
46 discriminating scenarios show which of these conditions it handles.

\section{Scenario Suite and Authoring}\label{sec:suite}

\begin{table}[b]
\caption{Scenario categories. \#: scenarios; D: discriminating scenarios; Literature: studies discussing
the history factors; ---: none in our survey; $^\ddagger$: the distinction between origin and direct BICs
is \tool{}'s own.}
\label{tab:suite}
\centering
\footnotesize
\setlength{\tabcolsep}{3pt}
\begin{tabular}{@{}l>{\raggedright\arraybackslash}p{0.53\columnwidth}rrl@{}}
\toprule
& Category & \# & D & Literature\\
\midrule
S1 & Basic semantic line changes & 15 & 2 & \cite{Sliwerski2005When,Kim2006Automatic}\\
S2 & Omission and deletion bugs & 17 & 2 & \cite{Sahal2018Additions,Rezk2021GhostCommit,Tang2024SemSZZ}\\
S3 & Intermediate and history noise & 41 & 3 & \cite{Kim2006Automatic,Williams2008SZZ,DaCosta2017Framework}\\
S4 & BFC-side noise and tangled commits & 17 & 2 & \cite{Kim2006Automatic,Dias2015Untangling,Tang2023NeuralSZZ}\\
S5 & Refactoring, move, rename after the BIC & 9 & 2 & \cite{CamposNeto2018RefactoringImpact,CamposNeto2019RevisitingSZZ}\\
S6 & Multiple BICs, duplicate/composite faults & 19 & 11 & \cite{Davies2014TextDependence,Lyu2024LinuxSZZ}\\
S7 & Context, message, non-local evidence & 7 & 2 & \cite{Tang2025LLM4SZZ,Shi2026AgenticSZZ}\\
S8 & Non-source, test-only, file-level cases & 8 & 2 & ---\\
\midrule
G1 & Refactoring, move, rename, rewrite & 15 & 2 & \cite{CamposNeto2018RefactoringImpact,Jodavi2022CodeTracker}\\
G2 & Clones, copies, origin vs.\ direct BIC$^\ddagger$ & 16 & 2 & \cite{Mondal2017CloneBugPropagation}\\
G3 & Multiple BICs and BFCs, partial fixes & 14 & 3 & \cite{Davies2014TextDependence,Park2012Supplementary,Lyu2024LinuxSZZ}\\
G4 & Branch, merge, cherry-pick, backport & 30 & 3 & \cite{Bludau2022PRSZZ,Lyu2024LinuxSZZ}\\
G5 & Rebase-specific history rewriting & 18 & 2 & \cite{Bird2009Perils,Lyu2024LinuxSZZ}\\
G6 & Revert-specific reintroduction & 11 & 2 & \cite{Shimagaki2016Reverts,Rosa2021DeveloperInformed}\\
G7 & Issue, message, non-local evidence & 18 & 2 & \cite{PerezRosero2024WIASZZ,Bludau2022PRSZZ,Shi2026AgenticSZZ}\\
G8 & Non-source, binary, generated, large files & 12 & 2 & ---\\
G9 & Evaluation policy variants & 12 & 2 & \cite{DaCosta2017Framework,Rosa2023Comprehensive}\\
\midrule
& Total & 279 & 46 &\\
\bottomrule
\end{tabular}
\end{table}

Table~\ref{tab:suite} lists the 17 categories: eight for classic scenarios (S1--S8, 133 scenarios) and
nine for graph scenarios (G1\nobreakdash--G9, 146). Each category collects history factors that the cited literature on
SZZ and on mining Git histories discusses.
S8 and G8 are \tool{}'s own: they cover faults and fixes outside ordinary source code, in configuration, test-only,
generated, binary, and large files, which the work we surveyed does not address explicitly. Oracles
name one BIC in 216 scenarios, two in 58, three in 3, and none in 2; 12 scenarios have two BFCs.
The scenarios are in Java, except seven in Python. Nothing in a repository marks it as a benchmark,
and commit metadata does not reveal the BICs: the best metadata baseline, the rule for one author,
passes 12 scenarios, and \cmd{message-admits-fault} passes none.

\textbf{Reliability of the scenarios.} To keep its scenarios reliable, \tool{} checks each one in the
three steps of the authoring flow (Fig.~\ref{fig:workflow}(a)): \cmd{validate --lint} checks the schema,
the references between commits and bugs, and lint rules, for example against messages that reveal a
commit's role; \cmd{generate} and \cmd{verify} build the repository and check it against the scenario,
including that the faulty statement appears at the commit declared to introduce it; and
\cmd{baselines} and \cmd{coverage} run on the generated repository and compare what they measure with
what the author declared, such as the tier. Every shipped scenario passes \cmd{validate --lint} and
\cmd{verify}. The Java code at each of the 2,180 commits that contain Java compiles with \cmd{javac},
except at three commits of one scenario, whose code needs a library that its Maven build declares.
For 84 scenarios, including 36 of the 46 discriminating ones, \tool{}'s tests also run the code at
each commit and check its output.

\textbf{Authoring support.} In a web editor, an author enters a source file's version at each commit,
marks the BICs and the BFC, and chooses the tier; the editor saves the scenario only after it passes
\cmd{validate --lint}. \cmd{new-scenario} and \cmd{new-graph-scenario} allocate the next free id and
write a skeleton scenario for the author to fill in.

\section{Demonstration and Preliminary Evaluation}\label{sec:eval}

After cloning SZZBench, four commands compare PyDriller's recorded predictions and the naive baselines
with the oracles:
\begin{lstlisting}
pip install -e .
szzbench generate-all-repos
szzbench evaluate --predictions \
  examples/pydriller-run/predictions.jsonl \
  --include-generated
szzbench baselines
\end{lstlisting}
\cmd{generate-all-repos} generates the 279 repositories in about three minutes, and \cmd{evaluate} takes
seconds: PyDriller's answer is correct (PASS) in 161 of the 279 scenarios, the result recorded in
\cmd{examples/pydriller-run/}.

\begin{table}[b]
\caption{Results of nine configurations and the naive baselines. Ctrl, Disc: passed control (of 233) and
discriminating (of 46) scenarios; P, R: precision and recall of the reported BICs, pooled over all scenarios.}
\label{tab:results}
\centering
\footnotesize
\setlength{\tabcolsep}{4pt}
\begin{tabular}{@{}lrrrrr@{}}
\toprule
Configuration & PASS & Ctrl & Disc & P & R\\
\midrule
PyDriller 2.12 & 161 & 160 & 1 & 0.781 & 0.689\\
PySZZ B-SZZ & 156 & 156 & 0 & 0.761 & 0.672\\
PySZZ AG-SZZ & 161 & 160 & 1 & 0.780 & 0.686\\
PySZZ MA-SZZ & 161 & 160 & 1 & 0.783 & 0.686\\
PySZZ L-SZZ & 153 & 153 & 0$^\dagger$ & 0.770 & 0.598\\
PySZZ R-SZZ & 156 & 156 & 0$^\dagger$ & 0.815 & 0.633\\
SZZ Unleashed, depth 3 & 28 & 24 & 4 & 0.518 & 0.771\\
SZZ Unleashed, depth 1 & 156 & 156 & 0 & 0.765 & 0.669\\
Token-level B-SZZ & 164 & 153 & 11 & 0.841 & 0.636\\
\midrule
\multicolumn{6}{@{}l}{\emph{Single baselines}}\\
\cmd{position-1} & 160 & 160 & 0 & & \\
\cmd{blame-deleted-lines} & 156 & 156 & 0 & & \\
\midrule
\multicolumn{6}{@{}l}{\emph{Upper bound: best baseline per scenario}}\\
Union of the 26 baselines & 233 & 233 & 0 & & \\
\bottomrule
\multicolumn{6}{@{}p{0.95\columnwidth}@{}}{\scriptsize $^\dagger$L-SZZ and R-SZZ report one BIC
per BFC, but every discriminating scenario requires two or more: their 0 reflects their output, not
how well they trace lines.}
\end{tabular}
\end{table}

\textbf{Setup.} We ran four SZZ tools in nine configurations through adapters that ship with \tool{}
(Table~\ref{tab:results}): the SZZ implementation of PyDriller~2.12~\cite{Spadini2018PyDriller};
the B-, AG-, MA-, L-, and R-SZZ variants of PySZZ~\cite{Rosa2021DeveloperInformed} as distributed with
the MV-SZZ replication package~\cite{Kondo2026MVSZZ}, with the library versions it pins; SZZ
Unleashed~\cite{Borg2019SZZUnleashed} at depth~1, which blames the deleted lines, and at its default
depth~3, which follows each blamed line two more steps back; and PySZZ's B-SZZ on histories that
cregit~\cite{German2019Cregit,Watanabe2024TokenSZZ} rewrites to one token per line, as in the MV-SZZ package.
No SZZ tool's code was changed, but three settings matter for reading the results: PySZZ runs as its own
configuration files set it up, with no issue-date filter (which drops candidates newer than the bug
report), except that R-SZZ reads every file rather than Java files only; SZZ Unleashed receives the
fix's commit date as the issue date, so every earlier candidate passes its date filter; and token-level
B-SZZ rewrites only Java files.
The results thus diagnose how each SZZ tool traces back from a known BFC under these settings; they
are not an empirical comparison of SZZ tools.

\textbf{Similar PASS counts, different scenarios.} PyDriller passes 161 scenarios and \cmd{position-1} 160,
but only 97 scenarios are passed by both: PyDriller passes 64 that the baseline fails, and the baseline
passes 63 that PyDriller fails. PyDriller, the five PySZZ variants, and SZZ Unleashed at depth~1 all pass
153 to 161 scenarios and share only 93 to 104 of them with the baseline. The number of passed scenarios
thus hides that these implementations and the baseline succeed on different scenarios; the per-scenario
results reveal it. Grouped by category, they also show where every SZZ tool fails: no configuration
passes any of the 17 scenarios of S2 (Table~\ref{tab:suite}), omission and deletion bugs, and in the 14
whose fix only adds code, none reports a BIC.

\textbf{A new representation moves the failure conditions.} Token-level B-SZZ passes 164 scenarios and
line-level B-SZZ (PySZZ B-SZZ) 156, but the net gain of 8 hides 31 gains and 23 losses. In SZZ-0136
(Fig.~\ref{fig:workflow}(b)), token-level blame reports both commits that wrote the faulty condition.
Token-level B-SZZ passes 11 of the 46 discriminating scenarios, including every one that another
configuration passes; none of the nine passes the other 35, which remain open. Eight of its gains are
in S3 (Table~\ref{tab:suite}), where formatting-only changes vanish at the token level. It also passes
6 of the 9 scenarios of S5, against at most 3 for any other configuration, but only 19 of the 30 of G4,
against 22 or 23 for PyDriller and the PySZZ variants. Conversely, one fix
turns \cmd{return locked;} into \cmd{return !locked;}: at the token level it only inserts \cmd{!} and
deletes nothing, so B-SZZ, which blames deleted lines, has nothing to blame. Token-level B-SZZ reports no
BIC in 63 scenarios, against 18 for line-level B-SZZ, and such insert-only fixes cause 14 of the 23
losses. A representation change thus trades one set of conditions for another, and \tool{} shows both.

\textbf{A setting explained.} SZZ Unleashed passes 156 scenarios at depth~1 but only
28 at its default depth~3, which has the highest recall of all configurations, 0.771. In all 133 scenarios
that pass at depth~1 but fail at depth~3, depth~3 reports every required BIC plus other commits, and
\cmd{exact\_set} fails a result with any extra commit. The drop is a
precision--recall trade-off rather than a defect.

\textbf{Environment and implementation.} The adapter writes a record for every scenario, and each
configuration reproduces its predictions when rerun with the same inputs, versions, and settings,
Python's hash seed included. Running the SZZ tools therefore also exposed problems that are not
algorithmic. PySZZ's replication package pins an old GitPython; with a
newer one, PySZZ logs a refused \cmd{git blame} option and carries on, so MA-, L-, and R-SZZ silently
change their answers in three scenarios. L-SZZ's PASS count also ranges from 152 to 155 with Python's
hash seed, so our runs fix the seed.

\textbf{Implications for MSR studies.} For studies that label data with SZZ, these runs suggest
reading the results by category rather than by PASS count, expecting no label from these SZZ tools for a
bug whose fix only adds code, and fixing and reporting library versions and hash seeds.

\section{Reuse and Limitations}\label{sec:reuse}

\textbf{Reuse.} (1) \emph{New SZZ tools:} an adapter runs the tool on each generated repository and
writes the records of Section~\ref{sec:design}; the four shipped ones have 300 to 850 lines of Python.
(2) \emph{Regression testing:} after changing an SZZ implementation, a maintainer runs
it again and compares the PASS or FAIL of every scenario with the results from before the change; the
scenarios that changed show what the change affected. (3) \emph{Diagnosing a variant or a setting:} as with the depth of SZZ
Unleashed, a researcher compares the per-scenario results with those of a base configuration and reads
the gains and losses per category. (4) \emph{New
conditions:} a researcher writes scenarios for a new condition, and the checks of
Section~\ref{sec:suite} test them. LLM- and agent-based methods take an SZZ tool's input, a repository and a
BFC, and can be evaluated wherever the inputs they need, such as issue text, are present in a scenario.
Natural extensions are scenarios in more languages and scenarios derived from fixes in real-project
oracles, which would tie the conditions to failures seen in practice.

\textbf{Limitations.} Synthetic histories cannot estimate an implementation's accuracy on real projects
or how often a factor occurs. Oracles define what a correct answer must report; they are not claims
about the causes of real bugs. The discriminating tier covers only scenarios whose oracle has several
BICs (Section~\ref{sec:floor}). The scenarios are mostly in Java, and
evidence outside Git, such as issue trackers and code review, is modeled only in part. Techniques that
locate BICs by running failing tests are out of scope. The editor handles only classic,
single-file, single-bug scenarios, and \tool{} has not been verified on Windows.

\section{Conclusion}\label{sec:conclusion}

An aggregate score cannot show why an SZZ implementation succeeds or fails; \tool{} shows it with
controlled Git histories, explicit oracles, and naive baselines. Its recorded runs reveal that
similar PASS counts hide different behavior and that a new representation moves failure conditions rather
than removing them. Other researchers can reproduce
these diagnoses and extend them with scenarios of their own.

\bibliographystyle{IEEEtran}
\bibliography{references}

\end{document}